\documentclass[conference,10pt]{IEEEtran}
\IEEEoverridecommandlockouts

\usepackage{textcomp}
\usepackage{gensymb}
\usepackage{multirow}
\usepackage[ruled,linesnumbered,noend]{algorithm2e}

\usepackage{cite}
\usepackage{subcaption}
\usepackage{amsmath,amssymb,amsfonts}
\DeclareMathOperator*{\argmax}{argmax} 
\usepackage[ruled,linesnumbered,noend]{algorithm2e} 
\usepackage{graphicx}
\usepackage{textcomp}

\usepackage{tabularx}

\usepackage{url}
\def\BibTeX{{\rm B\kern-.05em{\sc i\kern-.025em b}\kern-.08em
    T\kern-.1667em\lower.7ex\hbox{E}\kern-.125emX}}

\begin{document}

\title{MOTH- Mobility-induced Outages in THz: \\A Beyond 5G (B5G) application
\thanks{This project was funded by CMU Portugal Program: CMU/TMP/0013/2017- THz Communication for Beyond 5G Ultra-fast Networks.}
}

\author{\IEEEauthorblockN{Rohit Singh \textsuperscript{1}, Douglas Sicker \textsuperscript{1,2}, Kazi Mohammed Saidul Huq \textsuperscript{3} }
\IEEEauthorblockA{ \textsuperscript{1}  \textit{Engineering \& Public Policy}, \textit{Carnegie Mellon University}, Pittsburgh, USA \\
			       \textsuperscript{2}  \textit{School of Computer Science}, \textit{Carnegie Mellon University}, Pittsburgh, USA \\
			       \textsuperscript{3}   \textit{Instituto de Telecomunicações}, Aveiro, Portugal \\
Email: rohits1@andrew.cmu.edu, sicker@cmu.edu,kazi.saidul@av.it.pt}
}

\maketitle

\section*{ABSTRACT}
5G will enable the growing demand for Internet of Things (IoT), high-resolution video streaming, and low latency wire- less services. Demand for such services is expected to growth rapid, which will require a search for Beyond 5G technological advancements in wireless communi- cations. Part of these advancements is the need for additional spectrum, namely moving toward the terahertz (THz) range. To compensate for the high path loss in THz, narrow beamwidths are used to improve antenna gains. However, with narrow beamwidths, even minor fluctuations in device location (such as through body movement) can cause frequent link failures due to beam misalignment. In this paper, we provide a solution to these small-scale indoor movement that result in mobility-induced outages. Like a moth randomly flutters about, Mobility-induced Outages in THz (MOTH) can be ephemeral in nature and hard to avoid. To deal with MOTH we propose two methods to predict these outage scenarios: (i) Align-After-Failure (AAF), which predicts based on fixed time margins, and (ii) Align-Before-Failure (ABF), which learns the time margins through user mobility patterns. In this paper, two different online classifiers were used to train the ABF model to predicate if a mobility-induced outage is going to occur; thereby, significantly reducing the time spent in outage scenarios. Simulation results demonstrate a relationship between optimal beamwidth and human mobility patterns. Additionally, to cater to a future with dense deployment of Wireless Personal Area Network (WPAN), it is necessary that we have efficient deployment of resources (e.g., THz-APs). One solution is to maximize the user coverage for a single AP, which might be dependent on multiple parameters. We identify these parameters and observe their tradeoffs for improving user coverage through a single THz-AP.

\section*{Keywords}  Beyond 5G (B5G), Terahertz (THz), Indoor Mobility, Beam Alignment, Online Perceptron, Online Stochastic Gradient Descent (SGD).

\section{Introduction}
Various reports predict that in the next 5 years, wireless traffic will increase by 9-fold for on-demand and streaming games and by 12-fold for Virtual Reality (VR) and Augmented Reality (AR) services \cite{CVNI}. Given this trend, the demand for data rates will only increase in the near future. One of the key features of 5G is Ultra Reliable Low Latency Communications (URLLC) to support high-speed gaming, multimedia (stored and live), and yet-known killer applications \cite{ARQualcom}. The successful deployment of 5G will promote the densification of devices requiring an ultra-high-speed of $10Gbps$, a demand that will be hard for 5G technology to meet. We need fast and smart solutions to reduce latency, and support data rate demands Beyond 5G (B5G)  \cite{OurITS}. Fast encoding schemes \cite{JigSaw} and caching content close to the edge through Content Delivery Networks (CDNs) \cite{CDN} help but will not be enough in the future. Compressing and uncompressing videos results in unnecessary delays, which could have been used for transmitting more data. To transmit uncompressed video over-the-air requires more bandwidth; thus, we need more spectrum. The lower radio frequency bands ($<10GHz$) are almost saturated and require co-existence strategy \cite{MyTh} to be used for multimedia streaming. Moreover, the available bandwidth in the lower bands is one-order of magnitude less compared to the bandwidth available in higher frequency, such as mmWave, THz, infrared, and visible light. Finding additional bandwidth in the higher frequency is a viable option to be considered for future B5G applications.

The Terahertz (THz) band ($300 GHz -10THz$) is a promising area of additional bandwidth, possibly helping to push data rates up to $1Tbps$. The THz band is optimally located in the electromagnetic spectrum between the microwave and infrared bands \cite{OurTPRC}. In the future, THz band applications could be useful for VR, AR applications \cite{ARQualcom}, tactile internet, video streaming, healthcare, and public safety. Despite the tremendous potential, it also comes up with new challenges, which were not seen in the lower spectrum bands. Issues like molecular absorption, fast spread, and low penetration power can be challenging to deal with \cite{OurITS} \cite{Kazi1} \cite{Kazi2}. THz is quite sensitive to blockages, even humans \cite{HuBlk}, resulting in \textit{outages}. Thus, THz is envisioned to be used for short-range ultra-high data rate communication in an indoor environment, especially for Wireless Personal Area Networks (WPANs) \cite{IEEEwpan}. 

One of the promising solutions to gain higher throughput in THz band is the use of narrow beams to increase the antenna gain \cite{LiSteer} \cite{InferBeam}. Although narrower beamwidth results in higher antenna gains, it demands beam alignment between the access point (AP) and the user equipment (UE) \cite{OurGC}. Due to lower penetration power and beam misalignments resulting from the rapid mobility of the devices, it can cause connection outages, i.e., \textit{mobility-induced outages in THz (MOTH)}.  These small-scale outages are ephemeral in nature, similar to the random fluttering of a moth (an insect). Unlike an insect moth, MOTH can be destructive for wireless mobile communication. By modeling this small-scale user movement and device orientation \cite{OurGC}, we can avoid MOTH. An alternative solution to avoid MOTH is to increase the number of APs with overlapping coverage areas that can aid in decreasing the probability of these outages \cite{CovAchiIndoor}, but this increases power usage, handoff issues, waste of resources, and can still maintain uncertainty. Thus, it might be worth exploring methods that can improve user-coverage through a single AP and at the same time, avoid these mobility-induced outages. 

Currently, there is little research aimed at managing or mitigating these mobility-induced outages. Most of the existing research is focused on improving the average throughput through information-shower \cite{InfoShower}, identifying conservative frequency windows with minimal path loss \cite{LiBudTHz}, distance, and bandwidth adaptive methods \cite{DistBW}, or improving video streaming using local cache \cite{MManMMWave}. Analyzing mobility-induced outages for indoor users specifically in the THz band is critical to improving the quality of service (QoS) and throughput. 

Definition of mobility management changes as we enter the THz band. We are not only interested in the location and velocity of the user equipment, but also interested in the small-scale mobility, as the orientation of the device and the sudden oscillations on $x$, $y$ or $z$ axes. Some researchers have used user mobility data to consider tradeoffs between throughput and optimal beamwidths \cite{SmallScale}\cite{OurGC}, but it did not provide a solution or a framework to deal with this problem. There is ample research that proposes smart antennas to deal with interference and throughput related issues \cite{RandomMisAl} \cite{ContextBanditMisAl}, but most of these methods work on a batch model (i.e., train the model with pre-fetched data). Here, we propose online training models that learn the users' mobility patterns on-demand and predict outage scenarios. Although there is quite a bit of randomness associated with the user\textquotesingle s mobility when using AR and VR applications, we try to model these on a scenario-wise basis. It might be hard for machine learning algorithms to predict users' actions based on random action on all degrees of freedom; however, it is much easier when these movements are grouped into scenarios. In this paper, we propose that form of grouping. 

In this paper, we provide a solution to manage small-scale indoor mobility by reducing the time spent in the mobility-induced outages in THz (MOTH) \footnote{ Please note that physical blockages also result in outages. However, through MOTH, we try to identify outages that can result from small-scale mobility and narrow beamwidths in the THz, and provide solutions after that.}. MOTH uses state transition diagrams to model these small-scale mobility scenarios and propose a solution to mitigate these outages. Every time there is an outage the antennas at the AP and UE have to realign themselves, which is costly concerning time, i.e., the outage time (OT) and the realignment time (RT). For a single user scenario, it causes a decrease in throughput, while for a multi-user scenario, it also decreases user coverage. One solution is to make beam alignment as fast as possible \cite{FastmmWaveBA}, which is critical, but it does not reduce the outage time, and not what we are focusing on. Instead, we can try to predict these outage scenarios and avoid these time-consuming events. To do so, we explore two methods: Align-After-Failure (AAF) and Align-Before-Failure (ABF), which seek to predict these outages either through a hard margin specific to a system or a margin that can be learned by the system through observing the mobility patterns. Furthermore, we observe that for some fixed system parameters and mobility pattern, there are optimal beamwidths that can minimize the misalignment rate and maximize the user coverage. We show that beamwidths narrower than body rotation (yaw, pitch, and roll) may suffer from frequent misalignments if not appropriately monitored. Although increasing the number of APs will aid in improving the user coverage, to maximize the resource utilization, we analyze the case for a single AP and explore solution space. 

The rest of the paper is organized as follows. In section \ref{SysModel}, we discuss the system parameters that can maximize throughput by antenna gains and adaptive-frequency window selection and the mobility patterns associated with the human body in an indoor setting. By identifying these mobility patterns, we can reduce the need for beam alignments and outage time, either through hard time margins or dynamically learning these margins, as explained in section \ref{BAsch}. We evaluate our proposed methods in section \ref{Eval} and conclude in section \ref{Con}.

\setlength{\intextsep}{0pt}
\setlength\belowcaptionskip{-0.2 in}
\begin{figure}[t]
\centering
\includegraphics[width=3.5 in,height=2 in]{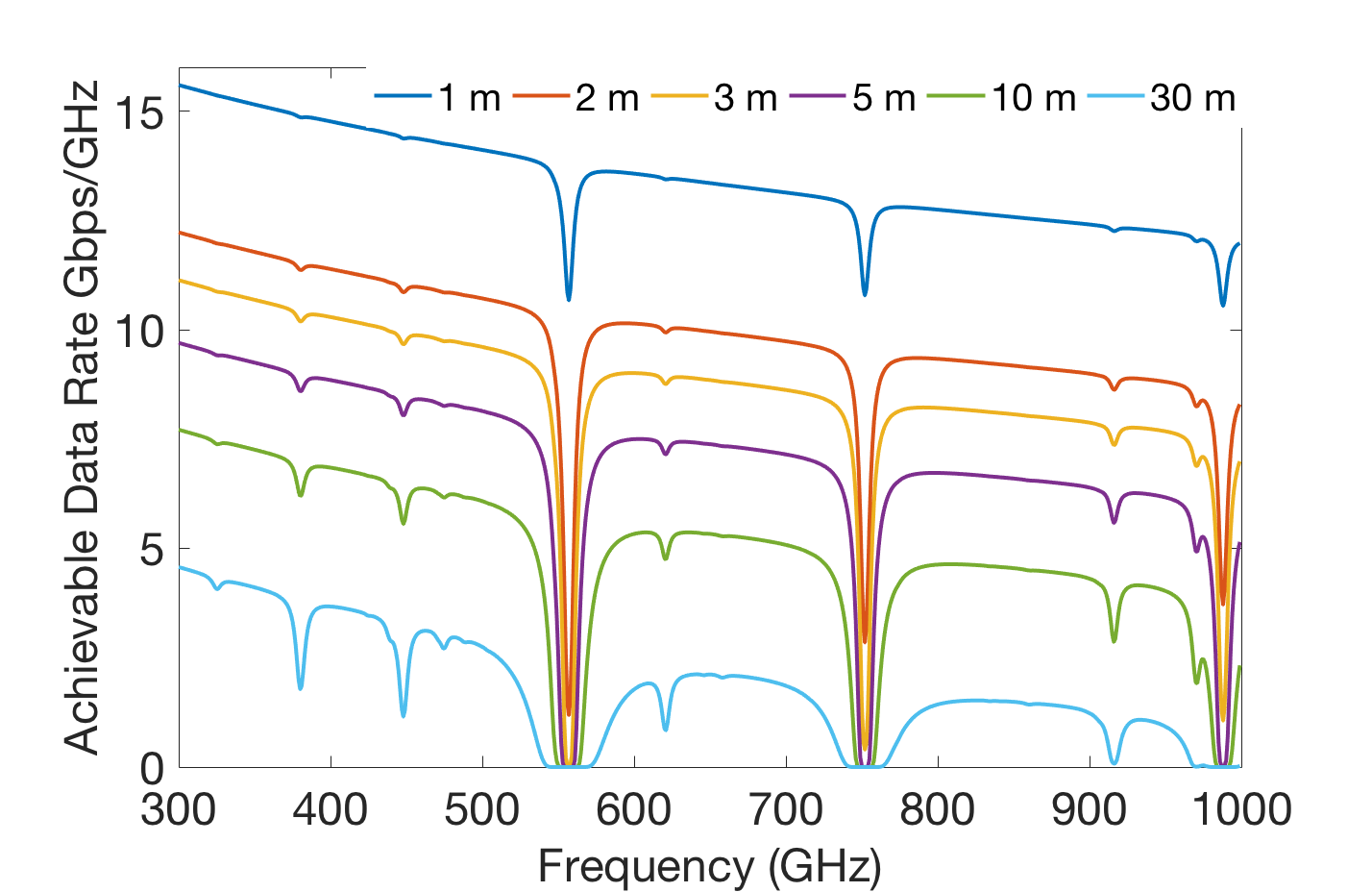}
\caption{Achievable data rate for different $d_j$ for a fixed relative humidity $\rho = 20\%$ and beamwidth $\delta=10 \degree$.}
\label{Adr10}
\end{figure}

    
\section{Models and Assumptions} \label{SysModel}

In this section, we briefly discuss the system models and assumptions that are applied in this paper.

\subsection{Link Budget Model}

One of the biggest challenges in THz communication is high absorption loss $L_A$ and spreading loss $L_S$, which are dependent on parameters, like the operating frequency $f$, water vapor concentrations $\rho$ \cite{ITUAtmAte} and the distance between the transmitter (TX) and receiver (RX) pair $d_j$ (Note: we drop the index for TX since we are analyzing only one AP for efficiency). The total path loss $L_T$ can be represented through Equation \ref{LossEq}, where $\mathcal{K}$ is the medium absorption coefficient dependent on the frequency range, relative humidity $\rho$, $f_c$ is the center frequency, $d$ is the seperation, and $c$ is the speed of light. For the rest of the paper, we have considered a $\rho$ of $5 g/m^3$ ($20\%$ humidity at room temperature of $25 \degree C$). To compensate for the high path loss, we need to increase the transmit power, which is still a challenge for small THz antennas. We keep transmit power fixed at $0 dBm$ \cite{LiBudTHz} and explore other workarounds, like antenna gain and bandwidth.

\abovedisplayskip=-4pt
\belowdisplayskip=4pt
\begin{eqnarray}
L_T=L_A (f,d,\rho)L_S (f,d)=e^{\mathcal{K}(f_c,\rho)d}((4 \pi d_j f_c)/c)^2    
\label{LossEq}
\end{eqnarray}

\setlength{\intextsep}{0pt}
\setlength\belowcaptionskip{-0.2 in}
\begin{figure}[t]
\centering
\includegraphics[width=3 in,height=2 in]{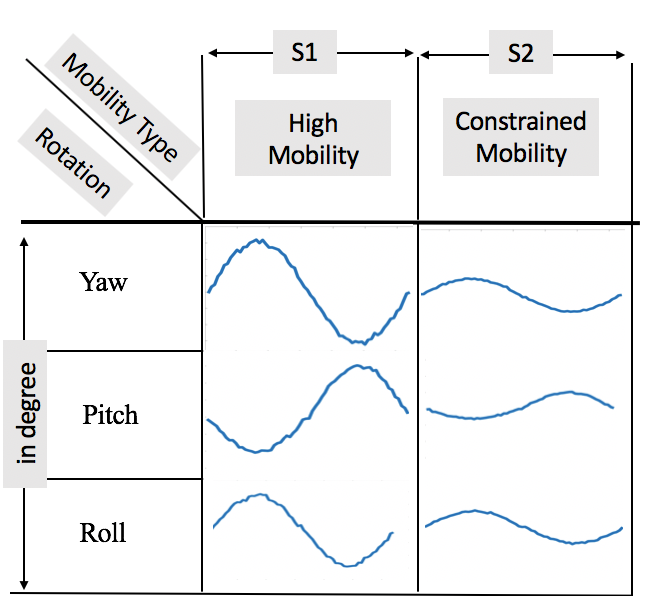}
\caption{Rate of change in yaw, pitch and roll for different human action types.}
\label{MobRot}
\end{figure}

The data rate $R_j$ for the user $j$ can be defined as $R_j=B_jlog(1+SNR_j)$, where $B_j$ is the bandwidth and $SNR_j$ is the signal-to-noise ratio \footnote{Please note that we did not consider signal-to-interference-noise ratio (SNIR), since in THz due to low penetration power and directional antennae interference is considered to be negligible.}. We can further calculate $SNR$ and represent $R_j$ as shown in Equation \ref{RateEq}, where $P_t$ is the transmit power of the TX, $G_t$ and $G_r $ are gains at that TX and RX respectively, and $N_o$ is the noise power spectral density of $-193.85 dB/Hz$. For simplicity, we assume an antenna pattern with perfect conical shaped main lobe \cite{SmallScale}, with the vertical and the horizontal beamwidth equal, i.e., $\delta_h= \delta_v=\delta$.

\abovedisplayskip=-4pt
\belowdisplayskip=4pt
\begin{eqnarray}
R_j=B_j log(1+\frac{P_t*G_t (\delta)*G_r (\delta)}{L_A (f_c,d_j )*L_S (f_c,d_j )* N_o*B_j })
\label{RateEq}
\end{eqnarray}

Based on Equation \ref{RateEq} , we can estimate the achievable data rate for different $d_j$ and center frequency $f_c$  as shown in Fig. \ref{Adr10}. Since the $L_T$ changes with every $GHz$ of frequency, which in turn changes the available bandwidth $B_j$ \footnote{The absorption loss in the THz spectrum is sensitive to multiple environmental factors, as shown in Fig. \ref{Adr10}. Thus, the \textit{frequency-windows} or the available bandwidth will change based on the path-loss $L_T$.}, we measure the achievable data rate in $Gbps/GHz$. Although $R_j$ scales with $\delta$, the sudden data rate dips become prominent as we increase the separation $d_j$. Thus, to obtain higher throughput, it is necessary that we have larger bandwidth, in other words, more continuous frequency-windows. We assume that an AP and UE can adaptively select a frequency-window, based on the parameters mentioned.

\subsection{Mobility Model} \label{MobiModel}

Despite the challenges associated with THz, using directional antennas with narrow beams can aid in communicating at ultra-high data rates. For narrow beamwidths, we need almost perfect alignment between the TX-RX antennas. These devices are highly sensitive to the orientation of the device and/or the object on which they are associated, i.e., fluctuations of hand, body, and head movement. Most of the THz applications, like VR and AR, will require fast body and hand movements; thus, compelling a user to move in all six degrees of freedom (DoF), with occasional body movements, such as stand, kick, duck, rotate, and dodge. Whether it is a high-intensity game or even a constrained indoor mobility scenario, such as walking or standing, narrow beam, THz communication links are susceptible to frequent misalignments. However, most of this movement follows a pattern. For example, in case of action games, particular movements are game-specific and are repeated quite often, or in case of regular walking, the body repeats a particular velocity and movement pattern. These features of human body movement can be parameterized for specific service types to provide the best QoS, namely High Mobility (e.g., high intense games) and Constrained Mobility (e.g., fast and slow walking).

Now that we have classified mobility patterns based on service types, it will be interesting to explore how the choice of beamwidths for each service type can affect the system throughput. For beamwidths wider than the rotational parameters of a particular service type, the location parameters will be prominent in identifying the occurrence of misalignments. While for relatively narrower beamwidths, misalignments will be sensitive to the rotation parameters. The yaw, pitch, and roll rotation represent a sinusoidal wave, where the rate is dependent on the user mobility type, as shown in Fig. \ref{MobRot} \cite{YawSin}. Fig. \ref{MobRot} implies that beamwidths narrower than body rotation may suffer from frequent misalignments if not appropriately monitored. A detailed mobility model is shown in \cite{OurGC}.

Please note a Time Division Duplexing (TDD) spectrum usage model is assumed, with a common channel for the users to communicate control signals with the AP.  In a TDD model, a user can occupy the whole channel for a portion of the time slot, which is called Channel Occupancy Time (COT).  We assume an adaptive COT value based on the user demand and distribution \cite{Part15++}, i.e., $COT_j^i=T_{slot}*R_j^*/R_j^i$, where $T_{slot}$ is the time slot for users to contest for a channel (or simulation time), $R_j^*$ is the requested data rate for user $j$, $R_j^i$ is the data rate available at time $i$. Since the system is blind to the mobility parameters of a user, the AP uses the ACK message to decide a successful communication and waits for a time $T_{fail}$. For a failed communication, the COT, along with the $T_{fail}$ can be termed as the outage time (OT). While after each failure, the time required to perfectly align the beams between a TX-RX can be termed as $T_{align}$ or the realignment time (RT).



\section{Beam Alignment Strategy} \label{BAsch}

In this section, we model the small-scale mobility scenarios into state transitions. We identify the successful states and the outage states, quantify the time spent in each state, and propose methods to minimize the time spent in the outage states. 
\setlength{\intextsep}{0pt}
\setlength\belowcaptionskip{-0.2 in}
\begin{figure}[t]
\centering
\includegraphics[width=3 in,height=2 in]{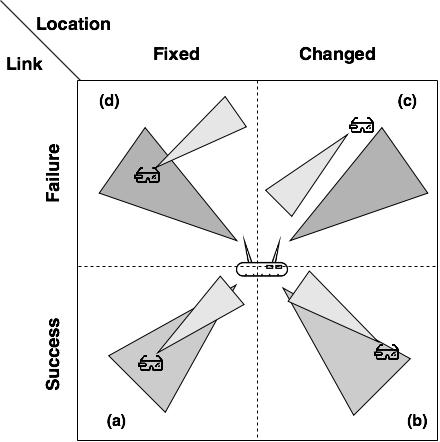}
\caption{Different Scenarios of Link Success and Failure.}
\label{AligEg}
\end{figure} 

\setlength\belowcaptionskip{-0.2 in}
\setlength{\abovecaptionskip}{0.2 in}
\begin{figure*}[t]
\centering
\begin{subfigure}[]{2.1 in}
\includegraphics[width=2.2 in,height=2 in]{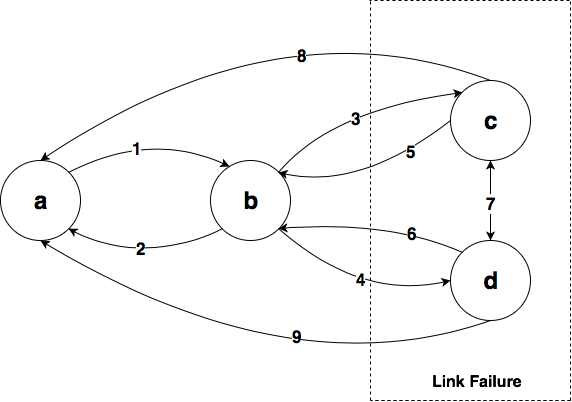}
\caption{Normal State Transition.}
\label{SD}
\end{subfigure}
~
\begin{subfigure}[]{2.1 in}
\includegraphics[width=2.2 in,height=2 in]{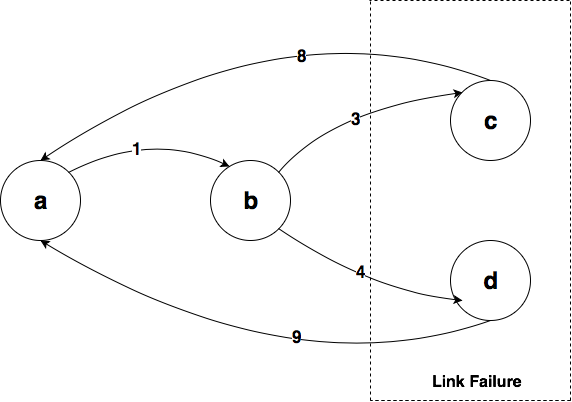}
\caption{AAF State Transition.}
\label{SD1}
\end{subfigure}
~
\begin{subfigure}[]{2.1 in}
\includegraphics[width=2.2 in,height=2 in]{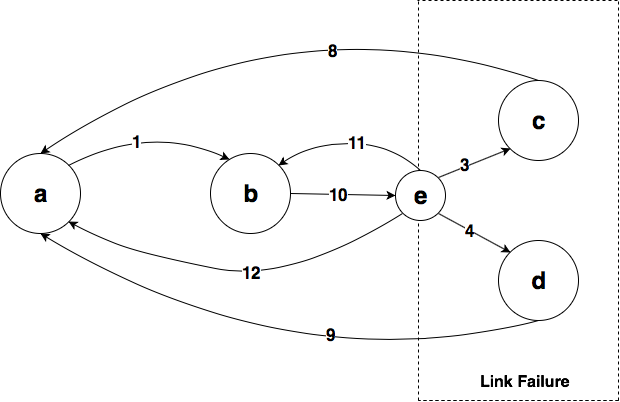}
\caption{ABF State Transition.}
\label{SD2}
\end{subfigure}
\caption{Link success-failure state diagram for a TX-RX pair (states shown in Fig. \ref{AligEg}).} 
\label{SDw}

\end{figure*}

\subsection{MOTH State Transitions}

Due to different user mobility patterns discussed in Section \ref{MobiModel}, the average throughput will vary. To avoid throughput drops in these scenarios, continuous beam alignment between the TX-RX pair is required, which consist of beam steering and beam training, i.e., the realignment time (RT). Time for beam alignment can grow as high as $100ms$ depending on the system and beam learning process. Let us assume a fast alignment of $T_{align} = 5ms$ \cite{FastmmWaveBA} and explore methods through which we can reduce this $100\%$ need for perfect beam alignment, as shown in part (a) of Fig. \ref{AligEg}.  

In a highly mobile scenario, the RT and OT will snowball as the number of users increases. Even small-scale orientational changes can cause outages. As shown in Fig \ref{AligEg}, we identify these small-scale mobile scenarios a TX-RX can encounter. We group these movements based on the success/failure of a link and the change in location/orientation of the UE relative to a fixed AP. Case (a) of Fig. \ref{AligEg} represents a \textit{perfect-alignment} between the TX and RX beams, which falls under a successful communication. However, in special scenarios, it might not be necessary always to have a perfectly aligned beam for successful communication, as shown in the case (b) of Fig. \ref{AligEg}. For example, we can still communicate with a misaligned link (named as \textit{partially-misaligned}) as long it can provide the requested data rate $R_j^*$ for a user $j$, as shown in part (b) of Fig. \ref{AligEg}. On the other hand, links might fail when the location ($X$, $Y$, and $Z$ coordinates) of a UE changes or if the orientation (yaw, pitch, and roll) has changed, as shown in cases (c) and (d) of Fig. \ref{AligEg} respectively. For example, for case (d) in Fig. \ref{AligEg} even if the user is within the AP's beam range, due to rotational motion of a user, it is a failed link.

Based on our four-case analysis shown in Fig. \ref{AligEg}, for a small change in time, TX-RX pairs relative locations and orientations can be bucketed into four states (a) to (d), resulting in a state transition diagram, as shown in Fig. \ref{SD}. In Fig. \ref{SD} state (a) is perfectly aligned beam between a TX-RX pair, while states (b), (c) and (d) are beam misalignments. State (b) is a partially-misaligned link subjected to the requested data rate $R_j^*$. Please note that for a heterogeneous data rate distribution, the state transition will be unique for each TX-RX pair, to be considered as a successful link. Similarly, from the explanation mentioned above, states (c) and (d) are considered as failure states. Thus, the transition between states (a) and (b) are considered as successful links, which does not involve any RT or OT, while states (c) and (d) are considered failed links, which costs RT. The transition between (b) and failed link sates are momentary based on relative orientation changes and costs OT. A beam alignment can trigger the transition from the state (b) to (a), which will cost RT. 

The time spent in each state, $\tau^s$ where s is the state is different and can significantly affect the system throughput. The users are provided with the best achievable data rate, $R_j^i$,  based on the environmental factors. At each state, we assume that the user occupies the channel, and some data is being transmitted, which cost $ COT_j^i $ time. For state (a), a perfect alignment state will cost $\tau^a= T_{align}+ COT_j^i $. Comparatively, state (b) can maintain connection without link failure (although with a reduced data rate of the bare-minimum data rate), which costs $ \tau^b =COT_j^i$ and does not need any extra time of $T_{align}$. On the other hand, states (c) and (d) are the mobility-induced outages, which costs $\tau^{c/d} =T_{fail}+ COT_j^i $. Since the users are free to move in all six degrees of freedom, a TX-RX pair can transition or loop around the state (b), (c) and (d), unless a perfect alignment is triggered. 

Our objective is to reduce the total time spent by a TX-RX pair transiting through these sates. To do so, we propose improvements over the normal transition diagram, i.e., Align-After-Failure (AAF) and Align-Before-Failure (ABF), shown in Fig. \ref{SD1} and \ref{SD2} respectively. In AAF, as shown in Fig. \ref{SD},  a perfect alignment (edge 2) is delayed until a link failure occurs, i.e., a transition to states (c) and/or (d). AAF drops edges 5 and 6, i.e., it does not allow transitions between state (b) and states (c)/(d) and triggers a transition to state (a) instead. AAF reduces significantly the oscillation between successful and failure states, which ought to happen in a small-scale mobile scenario. However, AAF can still go to failure state, causing unnecessary OT. A further improvement is to dynamically learn the mobility pattern of the user and make state transitions on-demand basis, as shown in Fig. \ref{SD2}. In ABF, we introduce a new learning state (e) that uses previous user mobility patterns to predict if the next user action will lead to a successful or failure link. The transition from the state (e) to either a successful or a failure state depends on the choice of the classifier, user mobility type, environmental factors, and data availability. We assume an online model where the state (e) receives user mobility data on-demand basis through a common-channel communicated between the TX and RXs.   

\subsection{Align After Failure (AAF)}
In AAF, the AP decides to perform a perfect beam alignment after a \textit{pseudo-failure} has occurred. A pseudo-failure is one where the AP waits for $T_ {fail}$ time for an ACK message to arrive before it infers that the link has failed. It is possible that in reality, the link has not failed, but this is a conservative choice that the algorithm makes to reduce the total time. The choice of $T_ {fail}$ can affect the efficiency of the method and can be computed based on the user service type and/or the system objective. For the sake of analysis, let us assume that the $T_{fail}=\alpha * T_ {slot}$, where $\alpha$ is a fraction and can be adjusted based on the system sensitivity.

A detailed pseudo-code for Method 1 for a single user is shown in Algorithm 1. Let us assume a method $Align()$ that performs perfect alignment between the TX-RX antennas, i.e., the transition to state (a), as explained above in the state diagrams. It returns a value of $\phi$, which is the relative transmission angle with the best achievable throughput. Let us assume that at time step $0$ the beams are perfectly aligned at angle $\phi^0$. For each time slot $i \in M$, where $M$ is the simulation time, Method 1 will check for a pseudo-failure. In Line \ref{Condi1}, the function $WaitForACK()$ is a timer to assure that the transmission does not take more than time $T_{fail}$. In case of successful communication, the previous transmission angle $\phi^{i-1}$ is retained; otherwise, a new beam alignment is triggered.

\setlength{\textfloatsep}{0pt}
\setlength{\intextsep}{0pt}
\begin{algorithm}[t]
{\small
  \Begin
  {
 	Set $\phi^0=Align()$ \\
	\For {$i \in {1,2, \cdot \cdot \cdot, M}$}	 
	{ 
		Transmit($\phi^i$) \\
		\If   {$WaitForACK()>T_{fail}$  \label{Condi1}}
		{
			$\phi^i=Align()$\\
		}
		\Else
		{
			$\phi^i= \phi^{i-1}$\\
		}
	}

 }	
\caption{Method 1- Align After Failure (AAF) for Single User $j$.}
\label{Algo1}
}
\end{algorithm}

\subsection{Align Before Failure (ABF)}

Although Method 1 can reduce the number of outage scenarios by considering a hard margin of $T_ {fail}$, it still costs more time and might make false predictions on link failures. Instead of assuming a hard margin, a better method will be to learn this margin dynamically based on user mobility patterns. This form of learning, i.e., online learning, improves its prediction with every new data set. Now, for the state (e), as shown in Fig. \ref{SD2}, we need a classifier algorithm that classifies between a partially-misaligned, i.e., state (b), and a mobility-induced outage  scenario, i.e., states (c) and (d). We consider two alternatives a single neuron online-perceptron model \cite{OnlineML} and an online stochastic gradient descent (SGD) \cite{OnlineML} to learn the classification threshold dynamically for each user individually. Although online-perceptron is faster, it draws a liner decision boundary which can often come up with the wrong prediction when trained with outlier data points. An improvement over this will be to uses online-SGD that draws a probabilistic decision boundary, but it is slow to train. Let us name the classifiers perceptron and SGD used in state (e) as Method 2 and Method 3, respectively. Please note that other state-of-the-art binary classification algorithms could be used to gain better results. In this paper, we merely are demonstrating the use of these algorithms to improve performance. To understand the relationship of user mobility, algorithm learning speed, and algorithm accuracy to make predictions adaptively, we consider these two forms of online classifiers.

As explained earlier, link failure can happen due to a change in location or orientation. It can also occur due to an object entering the path. Therefore, we need to train the classifier using user mobility parameters. Gain at both TX and RX, i.e., $ G_t $ and $G_r$, are directly related to the relative location. Additionally, we are not interested in the exact gain value or the exact location for a user at time instance $i$, but the amount of change in gains and data rate, i.e.,  $\Delta G_t,\Delta G_r, \Delta R$, compared to time instance $(i-1)$ that will trigger a need for beam alignment, i.e., a transition to state (a). Let us name these three features collectively as $Q^i_j$ for a user $j$ at time instance $i$. The choice of features is such that they linearly separate the data, one of the necessary conditions for a simple perceptron. Moreover, the transition in or out of state (b) is dependent on the requested data rate, which makes it necessary for the classifier to consider the data rate as a feature.

The objective of ABF is to reduce (i) the number of beam alignments required, and (ii) the time spent in failure states. Thus, let us define two sets of classes: ``$1$" need for beam alignment (or a probable transition to failure states) and ``$0$" otherwise (or a probable transition to successful states), represented by $y^i_j$ for a user $j$ at time instance $i$. Intuitively, the predicted label $\hat{y}$ $0$ or $1$ will result in edges $11$ and $12$, as shown in Fig. \ref{SD2}. A transition to state (b) means that beam alignment is not required, and the TX-RX can maintain the previously aligned angle $\phi^{i-1}$, or update it otherwise, i.e., $\phi_i=\textrm{Align()}$. 

Every service type has a pattern (introduced in Fig. \ref{MobRot}),  and with each new data point ($y^i_j$, $Q^i_j$) the classifier will try to draw a decision boundary that can separate the two classes $0$ and $1$. With every new data point $[{(y^0_j, Q^0_j),(y^1_j, Q^1_j), (y^2_j, Q^2_j), \cdot \cdot \cdot, (y^N_j, Q^N_j)}]$, the classifier will keep modifying the decision boundary by updating the gradients $\omega^i_k$ for every time instance $i$ and feature $k$ in set $Q_j$. We assume that a  bias term is already present in the set $\omega$, which corresponds to a set of 1\textquotesingle s appended in the set $Q_j$. In Method 2, the gradient update rule is based on the sign activation function, while in Method 3 the update rule will be probabilistic, with sigmoid activation function. Therefore, the predicted label $\hat{y}$ for Method 2 will be $\pmb{\textrm{sign}(\omega^T Q})$ and for Method 3 will be $\argmax_{\omega} p(y \mid Q)$, where $p(y \mid \pmb{Q})=1/(1+exp(-(\pmb{\omega^T Q})) $ is the conditional probability. Let $\pmb{\omega}$ and $\pmb{Q}$ denote vectors, and  assume a learning rate $\eta=0.01$. 

\setlength{\textfloatsep}{0pt}
\setlength{\intextsep}{0pt}
\begin{algorithm}[t]
{\small
  \Begin
  {
 	Set $\phi^0=Align()$, $net=LearnAlign(y^0,Q^0)$\\
	\For {$i \in {1,2, \cdot \cdot \cdot, M}$}	 
	{ 
		$\hat{y}=Predict(net, Q^i)$ \label{Train} \\

		\If   {$\hat{y}==1$}
		{
			$\phi_i=Align()$\\
			Transmit($\phi^i$)\\
		}
		\Else
		{
			Transmit($\phi^{i-1}$)\\
			$\phi_i= \phi^{i-1}$\\
			\If   {$WaitForACK()>COT^i$}  	
			{	$net=LearnAlign(1, Q^i)$ \label{ReTrain1} \\
			}
			\Else
			{	$net=LearnAlign(0, Q^i)$ \label{ReTrain2} \\
			}
		}
	}

} 	
\caption{Method 2 \& 3- Align Before Failure (ABF) for Single User $j$.}
\label{Algo2}
}
\vspace{-1mm}
\end{algorithm}

A generalized pseudo-code for both Methods 2 and 3 for a single user is shown in Algorithm 2. Assume a method $\textrm{LearnAlign}(y^i, Q^i)$ that learns the decision boundary measured by the feature parameters $Q^i$ and label $y^i$. Let the classifier used in the method $LearnAlign()$ emulate an online-perceptron and online SGD as explained above. At the initialization phase, we assume that the beam is perfectly-aligned, thus the label $y^0=1$. For a time instance $i$, let us consider a sample data point ($\hat{y}$, $Q^i$), where $\hat{y}$ is the predicted label shown in Line \ref{Train}. Based on the prediction made by the classifier $\hat{y}$, we decide if we need a perfect beam alignment or if we can still transmit in a partially misaligned angle $\phi^i $. If $\hat{y}==1$, then we align the beams and transmit at a new transmit angle $\phi^i$. Otherwise, for $\hat{y}==0$, we assume that the user is still partially-misaligned and transmit at angle $\phi^{i-1}$. In the case of a no alignment prediction, we retrain our model by checking for the true label based on the arrival of the ACK message. We retrain the model at Lines \ref{ReTrain1}, \ref{ReTrain2} with label $1$ if the ACK message arrives after $COT^i$ and $0$ otherwise, as true labels for $Q^i$.

\section{Evaluation} \label{Eval}

In this section, we evaluate the efficiency of our proposed methods and the classifiers. We then analyze how effectively MOTH is detected for a single and multi-user setting. 

\subsection{Environmental Setting}

For the simulations, we considered a cube room of $6 m$ with $N=40$ active users following a random waypoint model coupled with the mean and standard division of the location parameters mentioned in Table \ref{Tab1} \cite{RobotExercise}\cite{YawSin}. Each user can belong to either of the two service types, S1 or S2, with each UE requesting a minimum of $10 Gbps$ speed per time slot. The AP is placed at the middle of the ceiling, with adaptive frequency and bandwidth selection. Let us consider low transmit power devices of $0 dBm$ idle for WPAN.

Please note that to calculate the resultant $z$ coordinate for UEs, we add the $\Delta z$ oscillation over average human standing height of $1.5m$. To calculate the rotational changes, we use the peak values of the rotation parameters mentioned in Table \ref{Tab1} to replicate a sinusoidal wave pattern shown in Fig. \ref{MobRot}. Please note that while a body is in motion, the pitch and roll movements suffer through alternative deacceleration and acceleration, which might not present when a body is sedentary \cite{YawSin}. For a real emulation, we allow the rotational movements to vary by a small amount of noise.

\setlength{\textfloatsep}{0in}
\setlength{\intextsep}{0pt}
\setlength\belowcaptionskip{0.1 in}
\setlength\abovecaptionskip{0 in}
\begin{table}[t]  
\caption{User Mobility Parameters.}
\centering
\begin{tabular}{ |p{0.9cm}|p{0.9cm}|p{1.3cm}|p{1.5cm}|p{0.9cm}|p{0.9cm}|p{0.9cm}|} 
\hline
\multicolumn{2}{|c|}{} & $\Delta x$,$\Delta y$ &  $\Delta z$ & Yaw  & Pitch & Roll \\ \hline
\multicolumn{2}{|c|}{S1} & $1\pm0.5$ &    $0.5\pm0.05$	& $15.5 \degree$ & $13.8 \degree$ & $15 \degree$ \\ \hline
\multicolumn{2}{|c|}{S2}  & $0.9\pm0.7$  & $0.094\pm0.02$ & $4 \degree$ & $5  \degree$  & $5 \degree$ \\ \cline{2-7}
 \end{tabular}
\label{Tab1}
\end{table} 

\setlength{\textfloatsep}{-5in}
\setlength{\intextsep}{0pt}
\setlength\belowcaptionskip{0 in}
\begin{figure}[t]
\centering
\includegraphics[width=3.5 in,height=2.3 in]{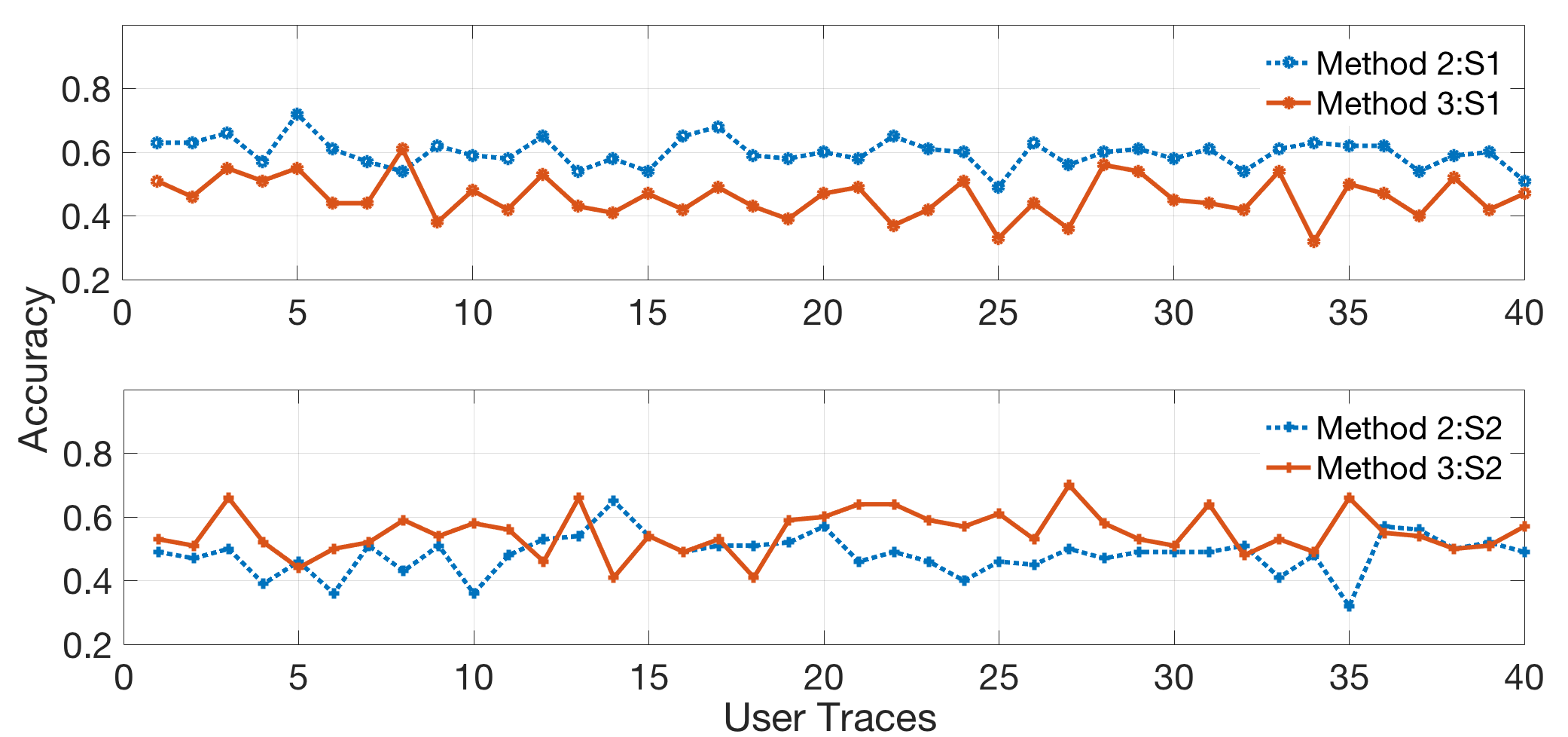}
\caption{Accuracy analysis for 40 different user traces.}
\label{AcUT}
\end{figure}

\setlength{\textfloatsep}{0pt}
\setlength\belowcaptionskip{-0.3 in}
\setlength{\abovecaptionskip}{0.1 in}
\begin{figure*}[t]
    \centering
    \setkeys{Gin}{width=\linewidth}
    
    \begin{tabularx}{\linewidth}{XXXX}

\includegraphics[width=3.5 in,height=2 in]{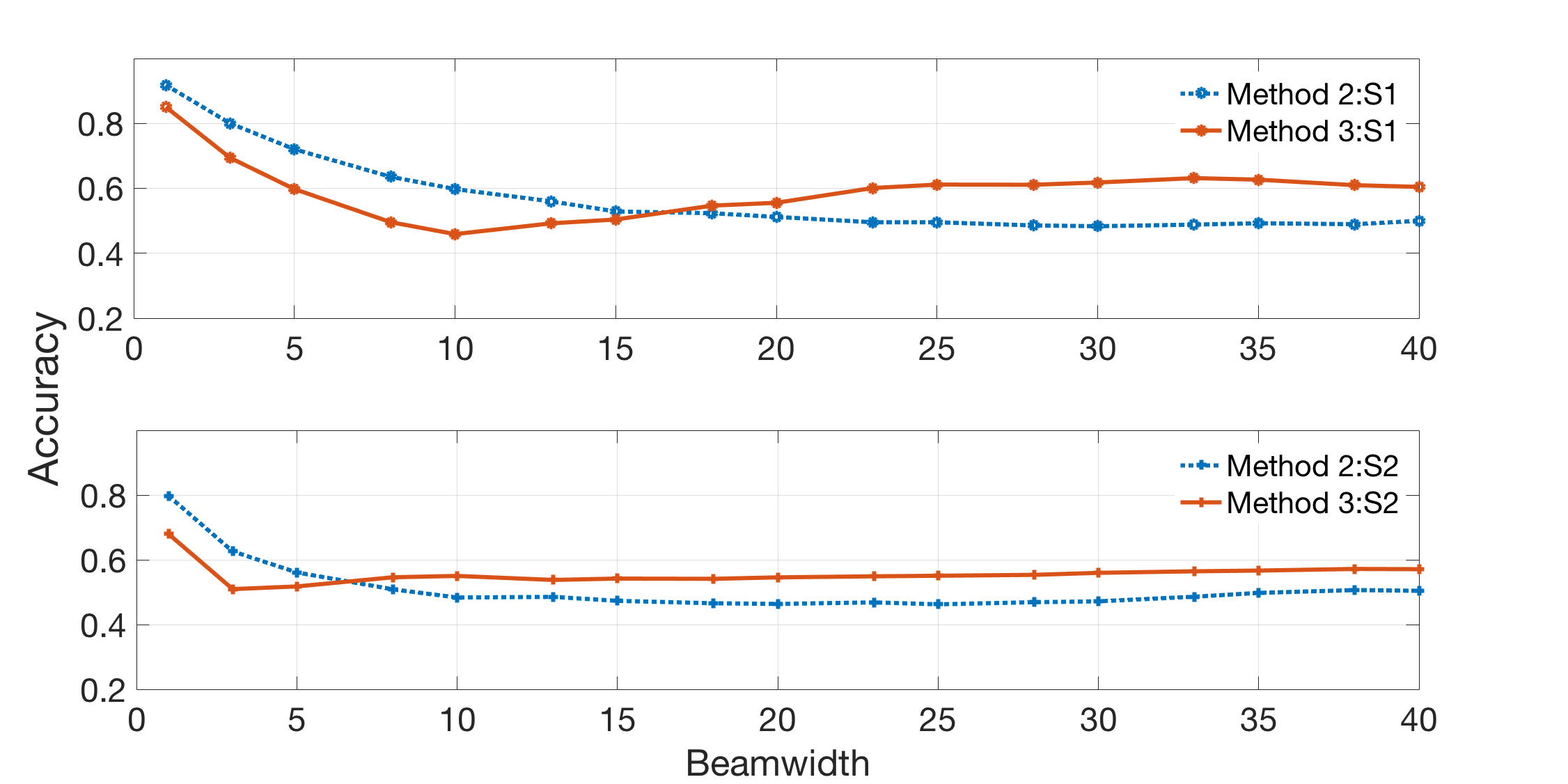}
\caption{Accuracy analysis for ABF.}
\label{Acu1}
&
\includegraphics[width=3.5 in,height=2 in]{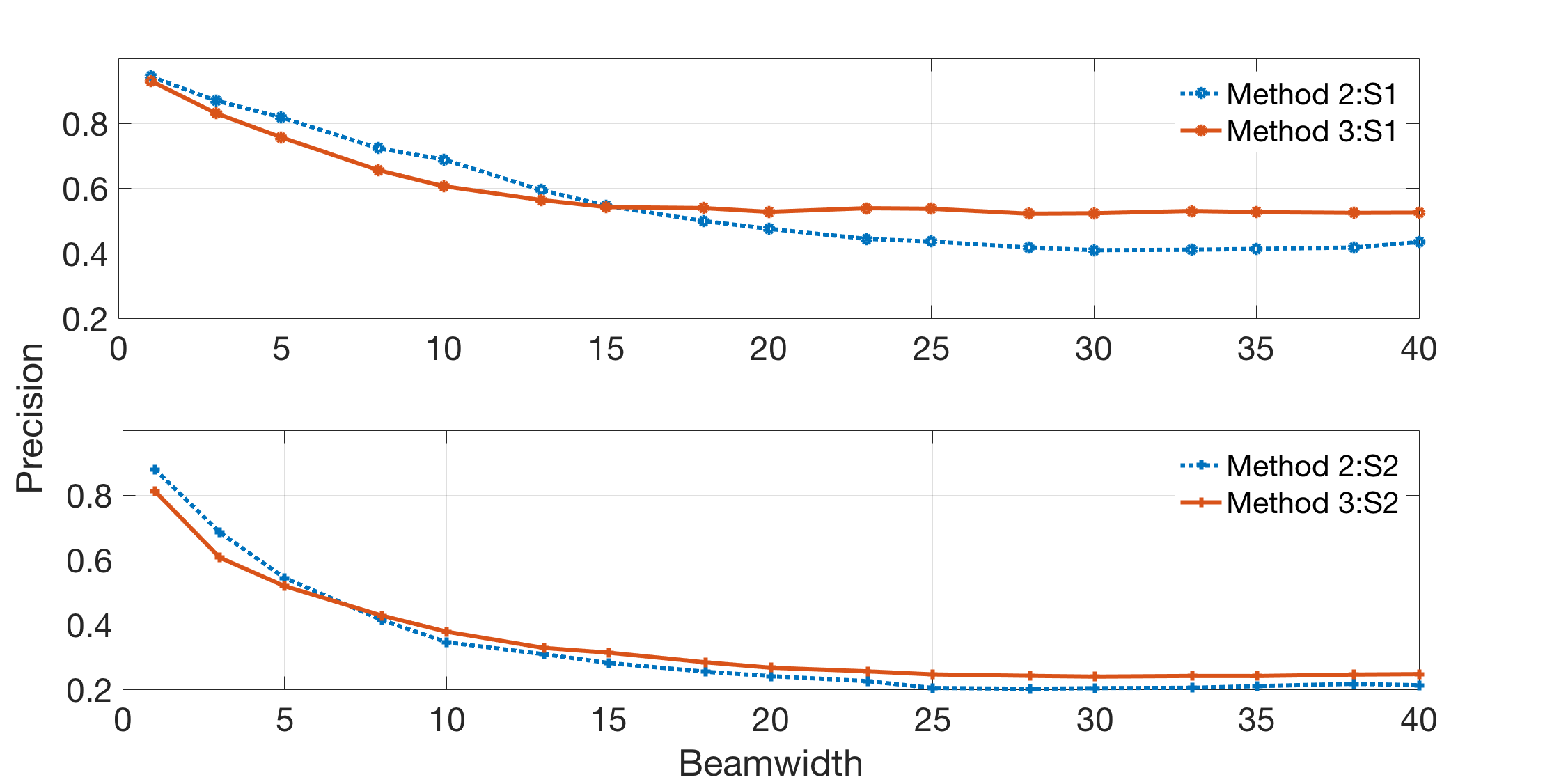}
\caption{Precision analysis for ABF.}
\label{Pre}
    \end{tabularx}
\end{figure*}

\setlength{\textfloatsep}{0pt}
\setlength\belowcaptionskip{-0.5 in}
\setlength{\abovecaptionskip}{0.1 in}
\begin{figure*}[t]
    \centering
    \setkeys{Gin}{width=\linewidth}
    
    \begin{tabularx}{\linewidth}{XXXX}

\includegraphics[width=3.5 in,height=2 in]{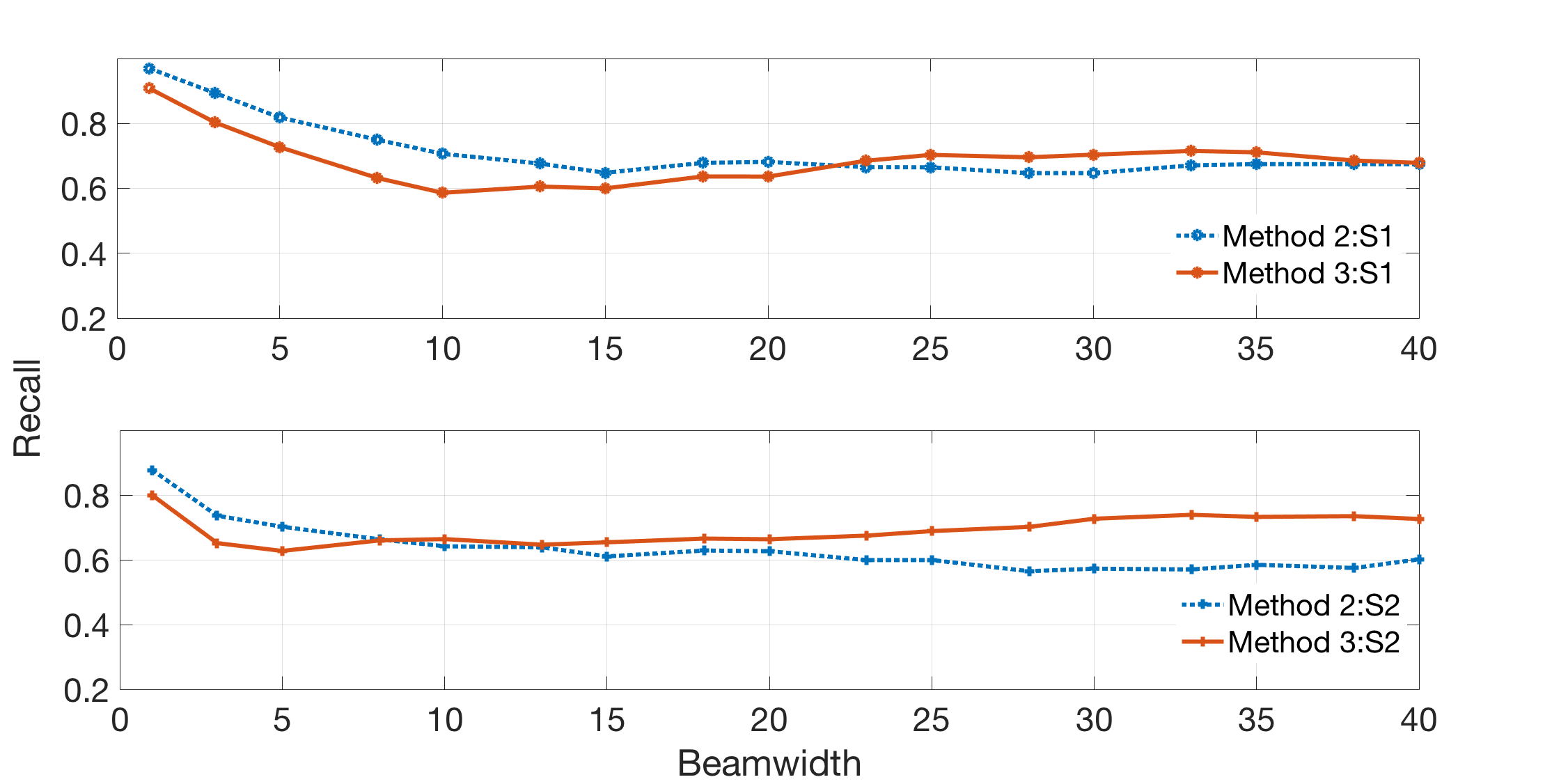}
\caption{Recall analysis for ABF.}
\label{Rec}
&
\includegraphics[width=3.5 in,height=2 in]{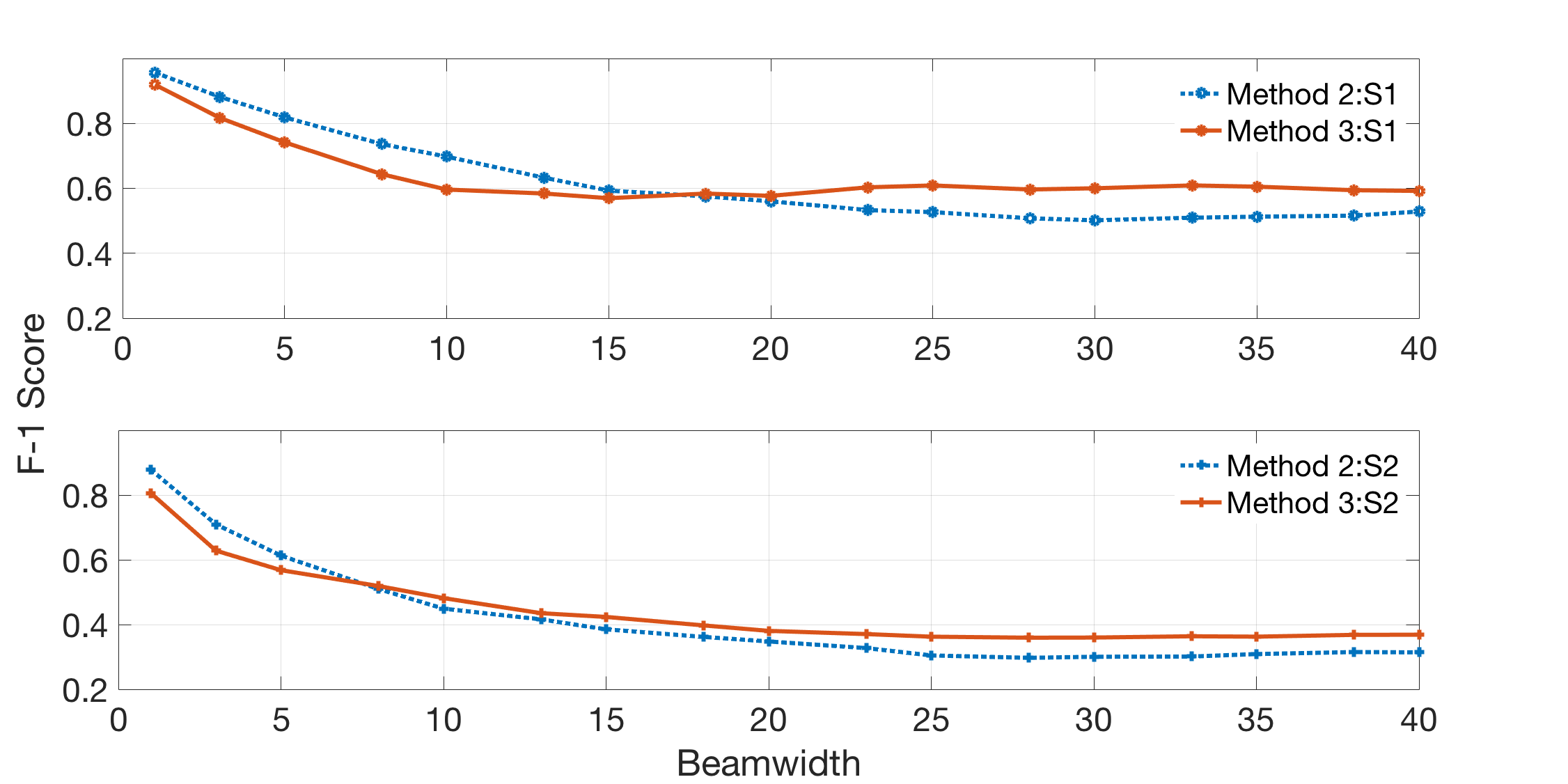}
\caption{F-1 score analysis for ABF.}
\label{F1}

    \end{tabularx}
\end{figure*}

\subsection{ABF Correctness Analysiss}

Before we move on comparing Methods 1-3 for a single and multi-user setting, let us analyze the accuracy of the prediction algorithms used in Method 2 and 3. To quantify the correctness of our model, we use the classical parameters such as accuracy, precision, recall, and F1 score \cite{OnlineML}.  

In our ABF model, true-positive (TP) is the correctly predicted beam alignment required, true-negative  (TN) is the correctly predicted no-beam alignment required, false-positive (FP) is the incorrectly predicted beam alignment required, and false-negative (FN) is the incorrectly predicted no-beam alignment required. Accuracy is the ratio between the correctly predicted observations to the total observations, i.e., $\frac{TP+TN}{TP+FP+FN+TN}$. We have considered $N$ different random human movement traces based on the parameters mentioned in Table \ref{Tab1}. For these $N$ different traces, the accuracy oscillates between $40\%$ to $80\%$, as shown in Fig. \ref{AcUT}, and variance is subjected to the method type and user movement pattern. For service type S1, Method 2 has higher accuracy compared to Method 3, while for service type S2 the accuracy almost overlays for both the methods. 

The result shown in Fig. \ref{AcUT} gives the impression that the prediction efficiency is likely dependent on the user mobility pattern. Furthermore, the antenna gain is subjected to the relative antenna angles and the antenna beamwidths $\delta$. On average, for all $N$ users, the accuracy stabilizes as the beamwidth increases, as shown in Fig. \ref{Acu1}. However, the correctness of the models changes with beamwidth for both service types.  For S1 users, which have high yaw, pitch and roll oscillations, Method 2 works better than Method 3 for beamwidths $\le15 \degree$ and the opposite for beamwidths less than $>15 \degree$. A similar effect is observed in S2, with a much lower pivot point. These pivot points are influenced by: (i) the rate of change of device orientation, and (ii) the chances of the user aligning with the AP antenna beam. For example, with a narrow antenna beamwidth and a higher rate of change of orientation, the user is likely to suffer mobility-induced outage and will require frequent beam alignments. A slow learning Method 3, which uses SGD, will take more time to correctly predict a need for beam alignment, while a quick learning perceptron will accurately identify these scenarios. 

Although accuracy is a good measure to check the correctness of a model, it is useful in case the cost of FP and FN are almost the same. However, in our case a FP can cost $\tau^{a}$ and a FN can cost $\tau^{c/d}$. Thus, it might be beneficial to look at precision, recall, and F1-score. Similar pivot points, as explained above, are observed in Fig. \ref{Pre}-\ref{F1}. On average, both methods show high recall and low precision. Low precision results in high FP, while high recall results in low FN. Since the user movements are quite random on all the 6-DoF planes, both methods become conservative and realign the beams most of the time, i.e., predict more FP over FN. In our simulation $\tau^{c/d}>\tau^{a}$, so we are better off with high recall values and low precision. To analyze the tradeoff between precision and recall for varying antenna beamwidths, we use the F1-score. The F1-score is the weighted average between precision and recall. Fig. \ref{F1} shows a high F-1 score above $50\%$ for service type S1 for both models and a relatively lower score of $30\%$ for service type S2. 

\subsection{Single User Analysis}

\begin{figure}[h]
\centering
\includegraphics[width=3.5 in,height=2.3 in]{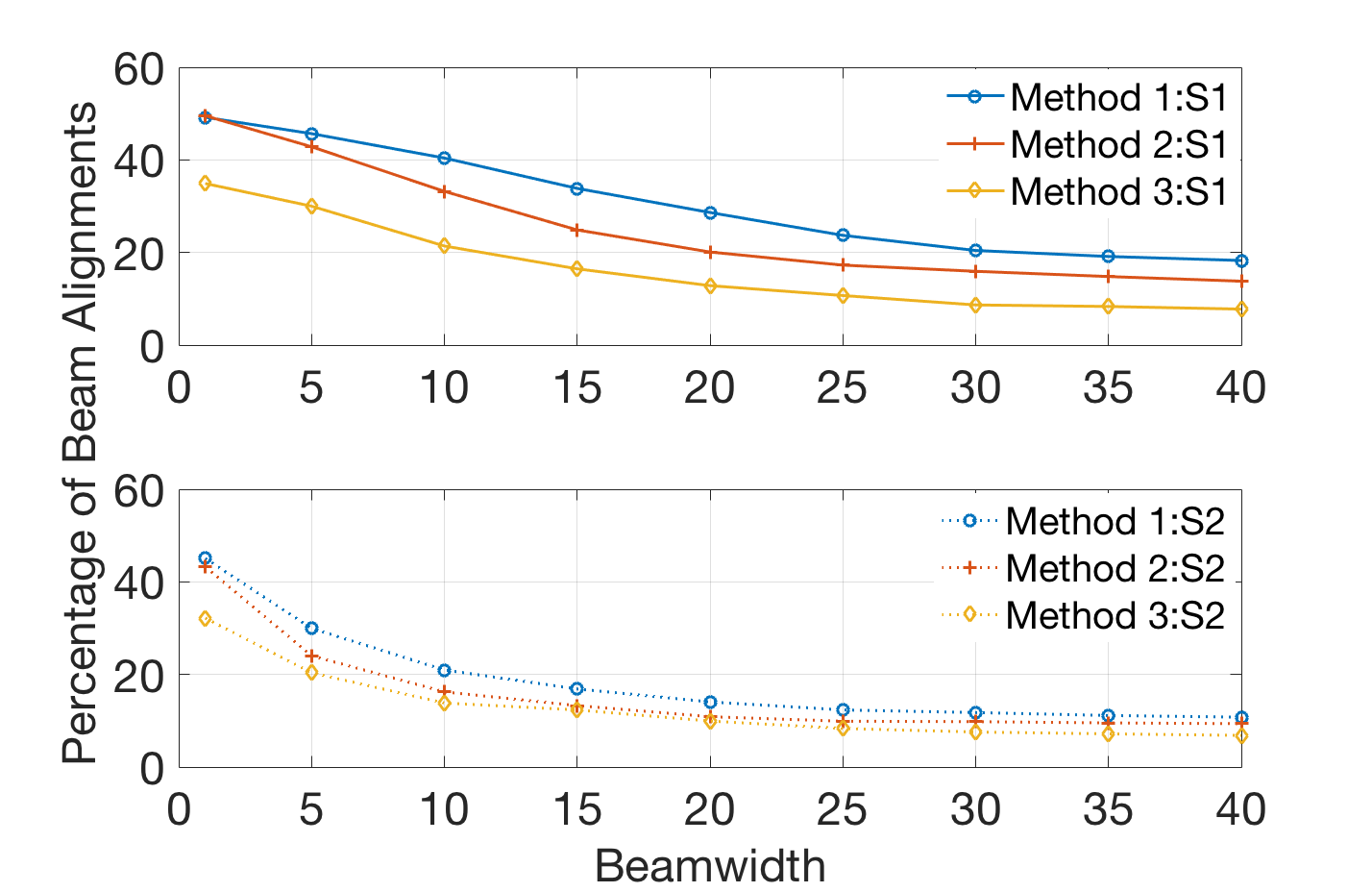}
\caption{Percentage of beam alignment required for different service types and antenna beamwidths.}
\label{U1M}
\end{figure}

Now that we have analyzed the correctness of the models let us analyze how the models allocate radio resources for a single-user and multi-user setting. Before we move on to a multi-user scenario, let us evaluate the average number of beam alignments required for a single user. Please note that a trivial case of perfect beam alignment will result in $100 \%$ alignment irrespective of beamwidth and/or service type. It can be observed from Fig. \ref{U1M} that compared to Method 1 (with $\alpha=0.25$), both Methods 2 and 3 can reduce the need for beam alignments significantly. Method 1 might not seem to be beneficial for service type S1, but for service type S2 (constrained mobility) all three methods show similar misalignment rates. Thus, Method 1 can be used in scenarios where the location is changing relatively linearly or if a user is relatively sedentary. Although the misalignment rate for both Methods 2 and 3 decreases with increase in beamwidth, there are significant gaps between the results based on service type.

\setlength{\intextsep}{0pt}
\setlength\belowcaptionskip{0 in}

\setlength{\textfloatsep}{0pt}
\setlength\belowcaptionskip{-0.3 in}
\setlength{\abovecaptionskip}{0.1 in}
\begin{figure*}[t]
    \centering
    \setkeys{Gin}{width=\linewidth}
    
    \begin{tabularx}{\linewidth}{XXXX}

\includegraphics[width=3.5 in,height=2 in]{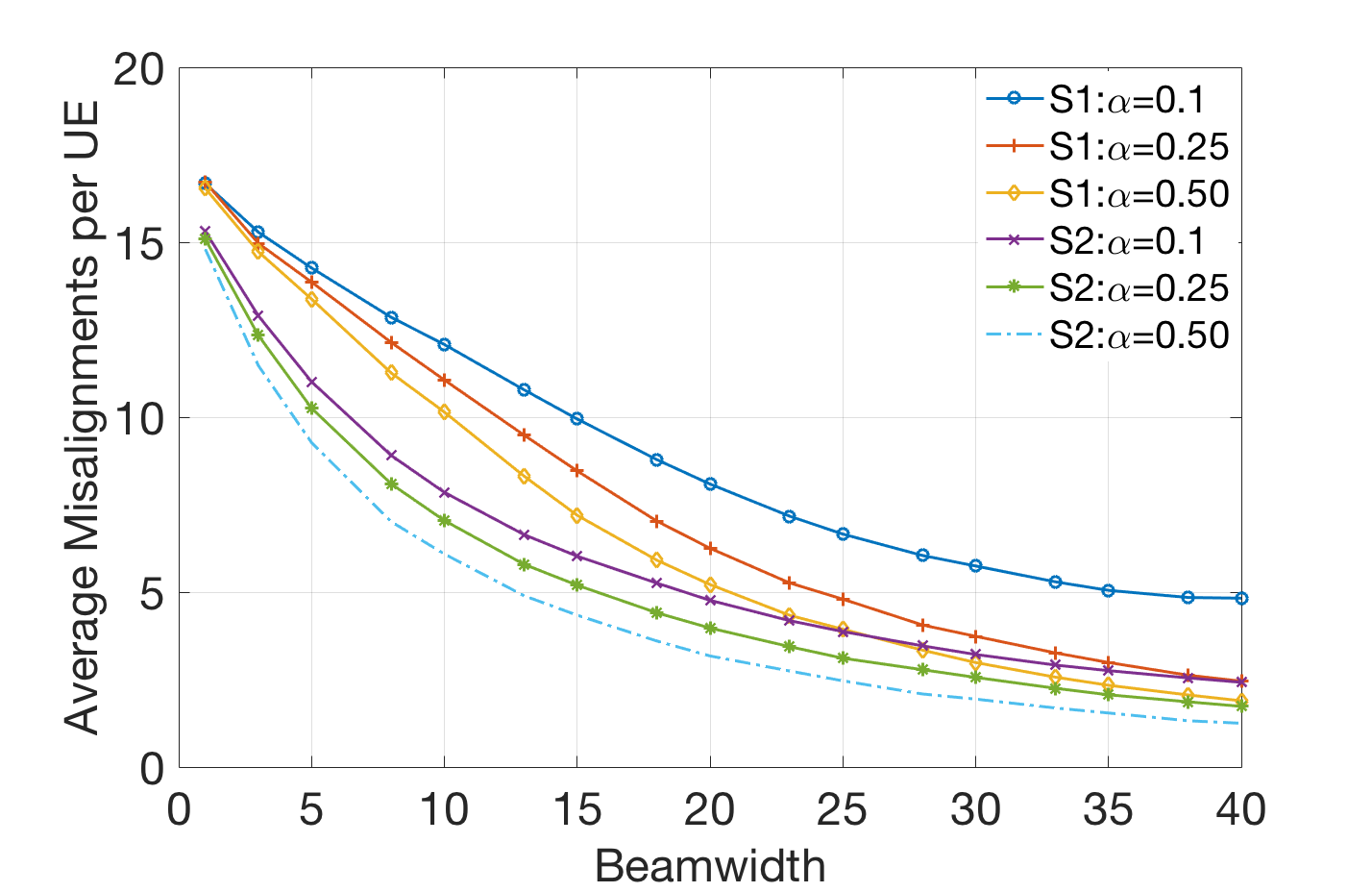}
\caption{AAF Misalignment.}
\label{UNM2}
&
\includegraphics[width=3.5 in,height=2 in]{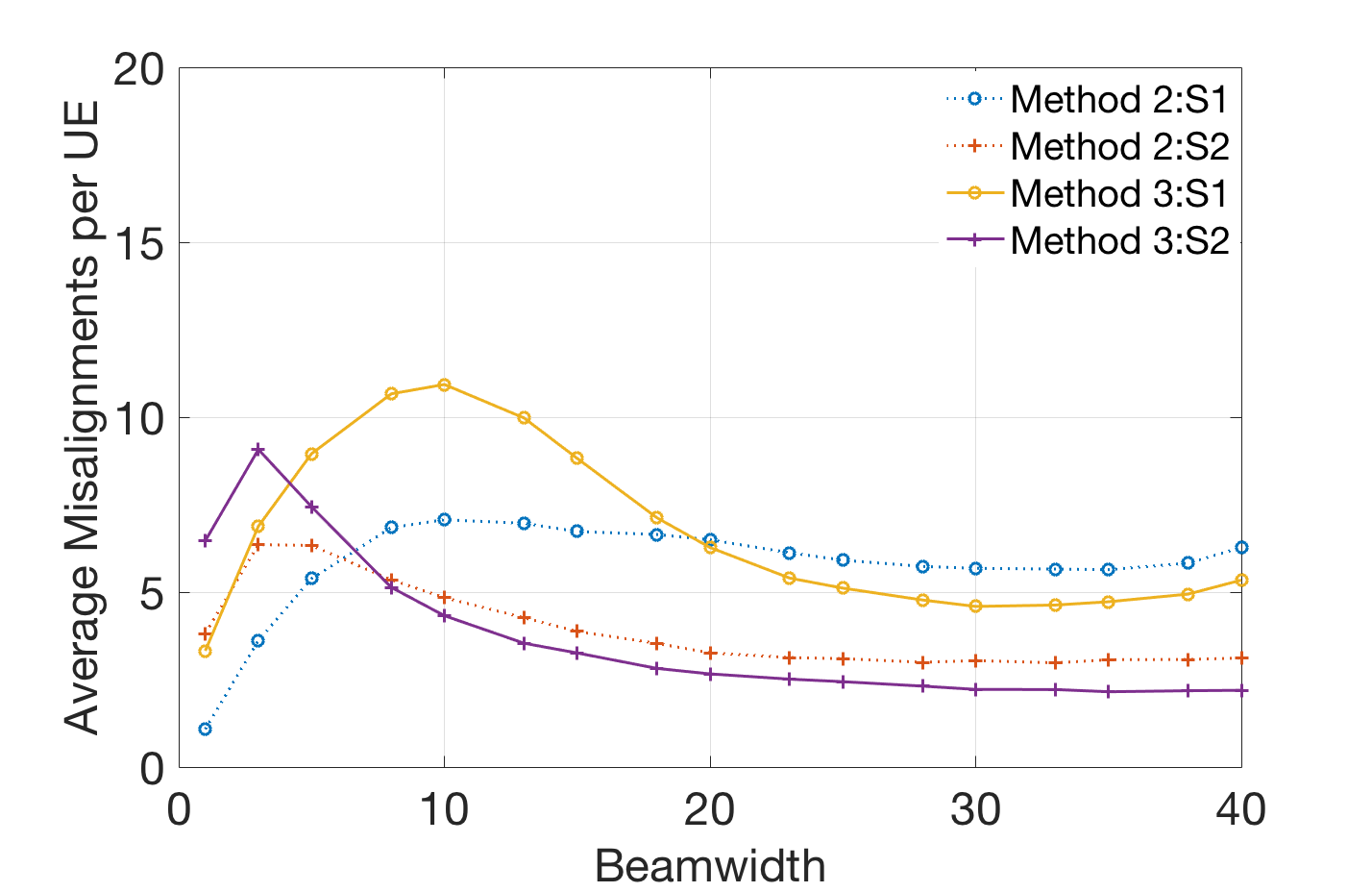}
\caption{ABF Misalignment.}
\label{UNM3}
    \end{tabularx}
\end{figure*}

\setlength{\textfloatsep}{0pt}
\setlength\belowcaptionskip{-0.5 in}
\setlength{\abovecaptionskip}{0.1 in}
\begin{figure*}[t]
    \centering
    \setkeys{Gin}{width=\linewidth}
    
    \begin{tabularx}{\linewidth}{XXXX}

\includegraphics[width=3.5 in,height=2 in]{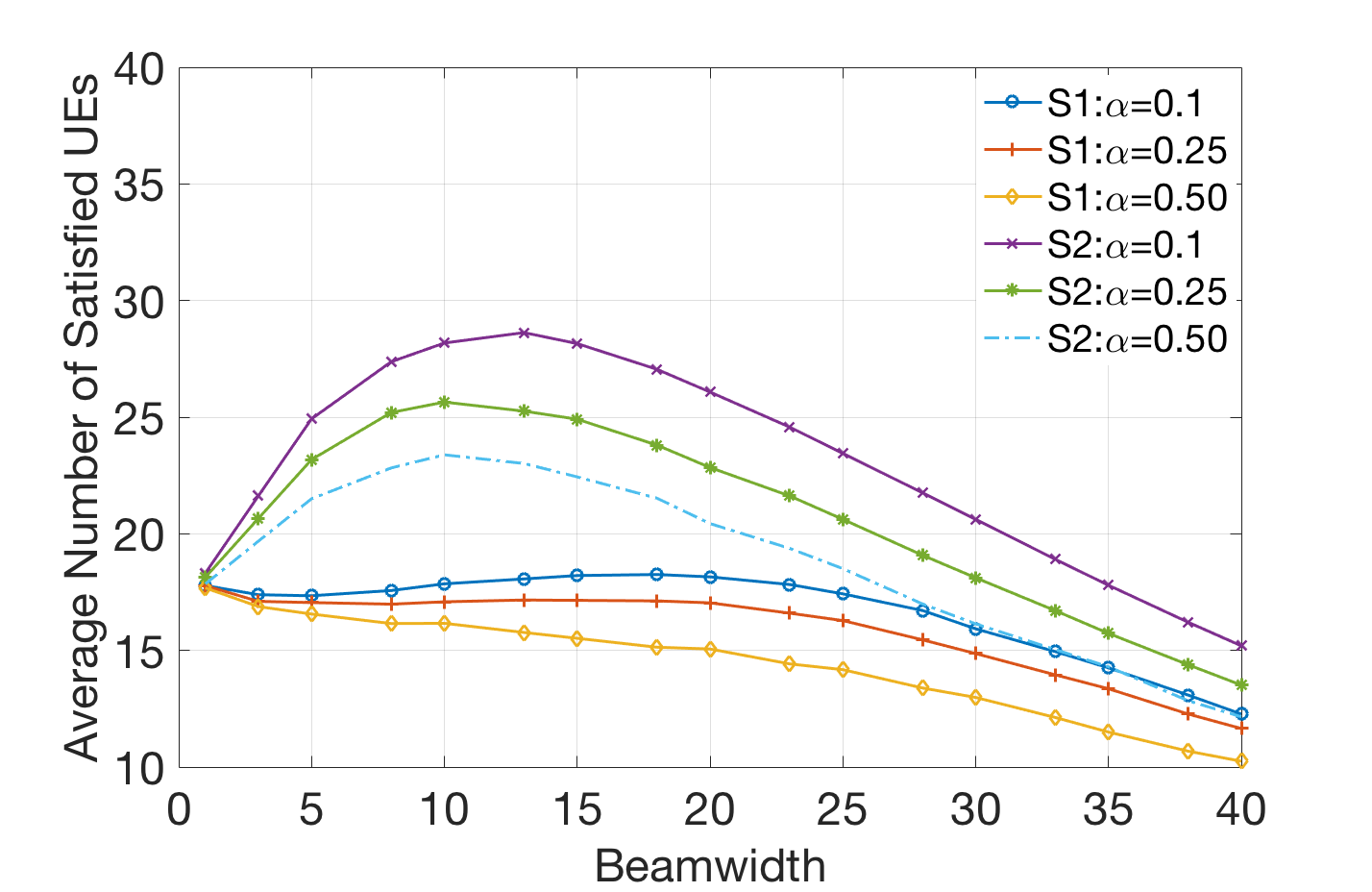}
\caption{AAF User Coverage.}
\label{UNU2}
&
\includegraphics[width=3.5 in,height=2 in]{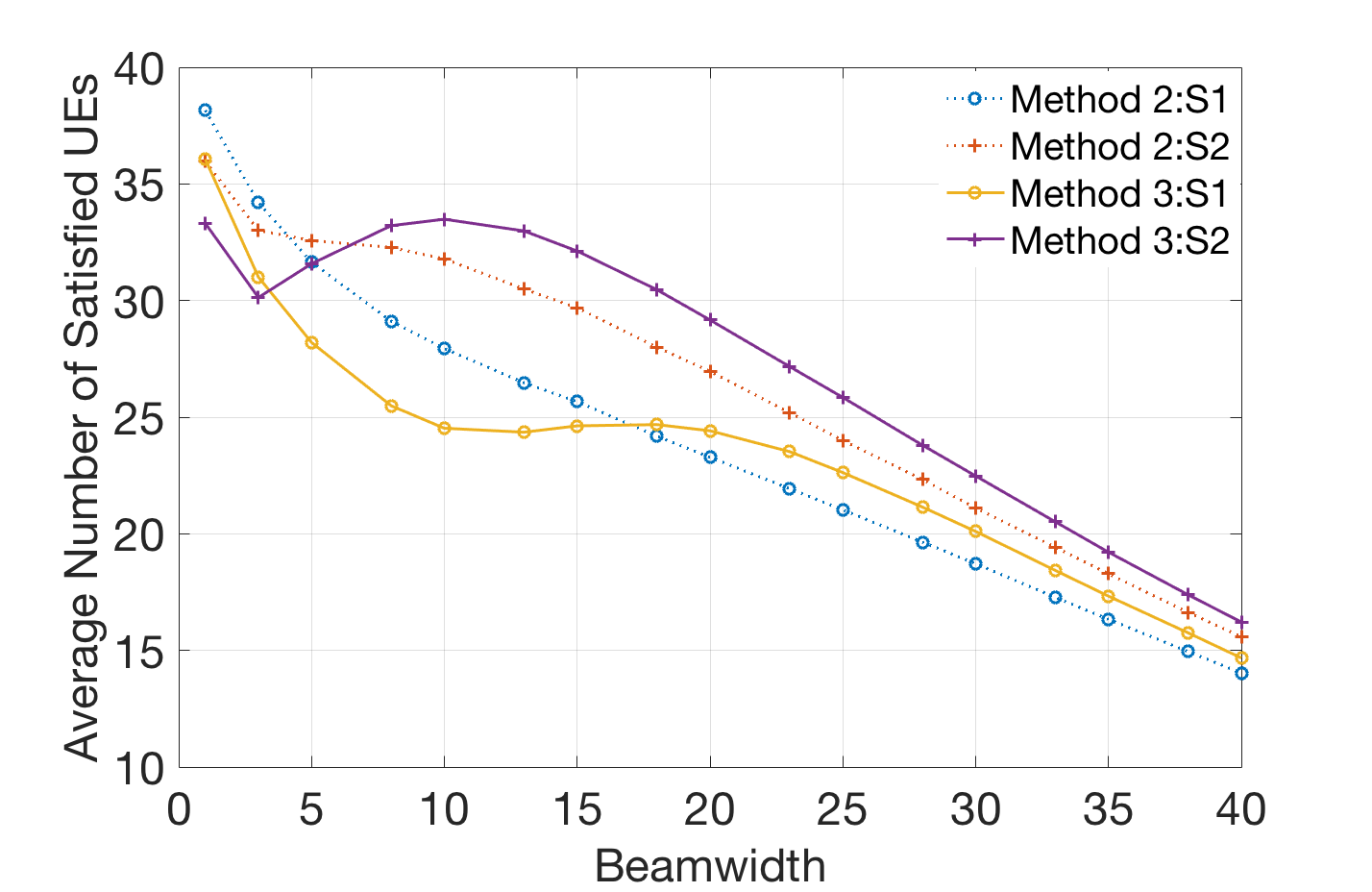}
\caption{ABF User Coverage.}
\label{UNU3}
    \end{tabularx}
\end{figure*}

\subsection{Multi User Analysis}

Fig. \ref{UNM2} - \ref{UNU3} shows the misalignment rates and user coverage for a multi-user setting for Methods 1, 2, and 3. Since for Method 1, the model is sensitive to the value $\alpha$, i.e., $T_ {fail}=\alpha * T_ {slot}$ we analyze the Method 1 separately. Although the misalignment rate for each user decreases with increase in beamwidths (shown in Fig. \ref{U1M}), for a multi-user setting, we observe user coverage peaks for certain $\delta$ and service types. At higher $\delta$, the antenna gains decrease and the $COT_j^i$ increases, thus reducing the user count. Despite higher gains at very low $\delta$, there are frequent misalignments due to yaw, pitch, and roll motions of the human body resulting in lower user coverage. Thus, we can observe that there are $\delta^{opt}_t, t \in{S1, S2}$ for all three methods. 

It is quite logical that misalignment rate for S1 (high mobility) will be greater than the rate for S2 (constrained mobility) due to the variability of the six DoF. As the $\delta$ value increases from $1 \degree$ to $40 \degree$, the average misalignment rate decreases rapidly for Method 1, as shown in Fig. \ref{UNM2}. The choice of $\alpha$ shows a significant effect in the case of S1. As the $\alpha$ value increase, the time occupied by a user $j$ at each time slot also increases, eventually decreasing the net user count. However, Method 1 cannot outperform Methods 2 and 3 shown in Fig. \ref{UNM3}. It seems when the yaw, pitch, and roll motions of the human body, as assumed in Table \ref{Tab1}, are comparable to the beamwidth, the probability of misalignment increases. Thus, Fig. \ref{UNM3} shows peaks that are dependent on the service types.

For both S1 and S2, Methods 2 and 3 shows significant improvement compared to Method 1 (shown in Fig.s \ref{UNU2} and \ref{UNU3}). For S1 at very low $\delta$, Method 2 performs much better than Method 3 because of the slow training rate of SGD. In Method 2, the online-perceptron quickly predicts the need for a beam alignment before a failure can occur. Although Method 3 performs better after $\delta^{opt} \ge 18 \degree$ the improvement is marginal. For S2 on an average Method 3 performs better than Method 2 for $\delta^{opt} \ge 5 \degree$. Method 3 in S2 does show a dip in user count for $\delta=3 \degree$, which can be due to the very small yaw, pitch and roll movements of the human body shown in Table \ref{Tab1}. Although for higher $\delta$ values, one can be indifferent between Methods 2 and 3, the choice of method might be sensitive as the APs and UEs scale up significantly. Please note that in our analysis, we assume the AP satisfies all UEs with the same optimal beamwidth; however, there can be user-specific beamwidths that can eventually result in both higher throughput and higher misalignment rate.


\section{Conclusion} \label{Con}

In this paper, we have proposed methods through which mobility-induced outages in THz (MOTH) can be reduced. It was interesting to observe that there exist optimal beamwidth values concerning the user service type and can be sensitive to the yaw, roll, and pitch of the body. We used the user orientations and different service types as a measure to predict an upcoming outage scenario. Both methods, Align-After-Failure and Align-Before-Failure, can show a significant decrease in outages; thus, increasing user coverage. Although ABF outperforms AAF in most cases, AAF can be used for static uses or constrained mobility scenarios. In ABF, the choice of the classifier will depend on system objective (throughput or user coverage), availability of resources (beamwidths), and environment (user service types).  Indeed, the proposed methods can be deployed scenario-wise based on optimal-beamwidth values. In future work, we will study user coverage and misalignment rates in a multi-user/service setting.


\begin{thebibliography}{51}
\providecommand{\natexlab}[1]{#1}
\providecommand{\url}[1]{{#1}}
\providecommand{\urlprefix}{URL }
\expandafter\ifx\csname urlstyle\endcsname\relax
  \providecommand{\doi}[1]{DOI~\discretionary{}{}{}#1}\else
  \providecommand{\doi}{DOI~\discretionary{}{}{}\begingroup
  \urlstyle{rm}\Url}\fi
\providecommand{\eprint}[2][]{\url{#2}}

\bibitem{CVNI} ``Cisco Visual Networking Index: Forecast and Trends, 2017-2022," Cisco, San Jose, CA, USA, Feb 2019. 


\bibitem{ARQualcom} ``Augmented and Virtual Reality: the First Wave of 5G Killer Apps," ABI research, Qualcomm, Jan 2017, 

\bibitem{OurITS} R. Singh, D. Sicker, ``Beyond 5G: THz Spectrum Futures and Implications for Wireless Communication," in  \textit{Proc. of 30th European Regional International Telecommunications Society (ITS) Conference}, 2019.

\bibitem{JigSaw} G. Baig, J. He, M. A. Qureshi, L. Qiu, G. Chen, P. Chen, Y. Hu, ``Jigsaw: Robust Live 4K Video Streaming," in \textit{Proc. of the 25th Annual International Conference on Mobile Computing and Networking (MobiCom)}, 2019. 

\bibitem{CDN} Y. Tang, X. Li, Y. Liu, C. Liu, Y. Xu, "Review of content distribution network architectures," in Proc. of 3rd International Conference on Computer Science and Network Technology, 2013. 

\bibitem{MyTh} R. Singh,  "Spectrum Sharing Opportunity for LTE and Aircraft Radar in the 4.2 - 4.4 GHz Band," M.S. thesis, Dept. Comput. Sci., Illinois Institute of Tech., Chicago, USA, 2017.

\bibitem{OurTPRC} R. Singh, W. Lehr, D. Sicker, K. M. S. Huq, ``Beyond 5G: The Role of THz Spectrum," in \textit{Proc. of the 47th Research Conference on Communication, Information and Internet Policy (TPRC),} 2019

\bibitem{Kazi1} S. A. Busari, K. M. S. Huq, S. Mumtaz, J. Rodriguez, ``Terahertz Massive MIMO for Beyond-5G Wireless Communication," in \textit{Proc. of IEEE International Conference on Communications (ICC),} 2019.

\bibitem{Kazi2} K. M. S. Huq, et.al.,``Terahertz-Enabled Wireless System for Beyond-5G Ultra-Fast Networks: A Brief Survey," \textit{IEEE Network}, vol. 33, no. 4, pp. 89-95

\bibitem{LiSteer} M. K. Haider, Y. Ghasempour, D. Koutsonikolas, E. W. Knightly, ``LiSteer: mmWave Beam Acquisition and Steering by Tracking Indicator LEDs on Wireless APs," in  \textit{Proc. of the 24th Annual International Conference on Mobile Computing and Networking (MobiCom)}, Oct. 2018, pp. 273-288, 

\bibitem{InferBeam} Q. Z. Sai, H.T. Kung, Y. Gwon, ``InferBeam: A Fast Beam Alignment Protocol for Millimeter-wave Networking," Feb 2018.  Available: \url{https://arxiv.org/pdf/1802.03373.pdf}.

\bibitem{OurGC} R. Singh, D. Sicker, ``Parameter Modeling for Small-Scale Mobility in Indoor THz Communication," in \textit{Proc. of IEEE Global Telecommunications Conference (GLOBECOM),} 2019. 

\bibitem{CovAchiIndoor} A. Moldovan, P. Karunakaran, I. F. Akyildiz, W. H. Gerstacker, ``Coverage and achievable rate analysis for indoor terahertz wireless networks,"  in  \textit{Proc. of IEEE International Conference on Communications (ICC)}, May 2017. DOI: 10.1109/ICC.2017.7996402

\bibitem{HuBlk} B. A. Bilgin, H. Ramezani, O. B. Akan, ``Human Blockage Model for Indoor Terahertz Band Communication," in \textit{Proc. of IEEE International Conference on Communications Workshops (ICC Workshops),} 2019. 

\bibitem{IEEEwpan} IEEE 802.15 WPAN\texttrademark -Terahertz Interest Group (IGthz). Avialble:\url{http://www.ieee802.org/15/pub/TG3d/index_IGthz.html}

\bibitem{InfoShower} V. Petrov, D. Moltchanov, Y. Koucheryavy, ``Applicability assessment of terahertz information showers for next-generation wireless networks," in  \textit{IEEE International Conference on Communications (ICC)}, May 2016. 

\bibitem{LiBudTHz} T. Schneider, et.al., `` Link Budget Analysis for Terahertz Fixed Wireless Links,"  \textit{IEEE Transactions on Terahertz Science and Technology},  vol. 2, no. 2, pp. 250-256, 2012. 

\bibitem{DistBW} A. A. Boulogeorgos ; E. N. Papasotiriou ; A. Alexiou, ``A Distance and Bandwidth Dependent Adaptive Modulation Scheme for THz Communications," in  \textit{Proc. of IEEE 19th International Workshop on Signal Processing Advances in Wireless Communications (SPAWC)}, Jun. 2018. 

\bibitem{MManMMWave} O. Semiari, W. Saad, M. Bennis; B. Maham, ``Mobility Management for Heterogeneous Networks: Leveraging Millimeter Wave for Seamless Handover,"  in \textit{Proc. of IEEE Global Communications Conference (GLOBECOM)}, Dec. 2017. 

\bibitem{SmallScale} V. Petrov, D. Moltchanov, Y. Koucheryavy, J. M. Jornet, ``The effect of small-scale mobility on terahertz band communications," in  \textit{Proc. of the 5th ACM International Conference on Nanoscale Computing and Communication (NANOCOM)}, no. 40, Sept. 2018. 

\bibitem{RandomMisAl} S. Priebe, M. Jacob, T. Kürner, ``Affection of THz indoor communication links by antenna misalignment," in  \textit{Proc. of the 6th European Conference on Antennas and Propagation (EUCAP)}, Mar. 2012. 

\bibitem{ContextBanditMisAl} M. Hashemi, A. Sabharwal, C. E. Koksal, N. B. Shroff, ``Efficient Beam Alignment in Millimeter Wave Systems Using Contextual Bandits," in  \textit{Proc. of the IEEE International Conference on Computer Communications (INFOCOM)}, Apr. 2018. 

\bibitem{FastmmWaveBA} H. Hassanieh, et.al., ``Fast millimeter wave beam alignment," in  \textit{Proc. of the ACM Special Interest Group on Data Communication (SIGCOMM)}, Aug 2018, pp. 432-445, 

\bibitem{YawSin} T. Imai, S. T. Moore, T. Raphan, B. Cohen, ``Interaction of the body, head, and eyes during walking and turning," \textit{ Exp. Brain Res., Springer}, vol. 136, no. 1, pp. 1-18, 2001.  

\bibitem{RobotExercise}T. Otani, et.al, ``Upper-Body Control and Mechanism of Humanoids to Compensate for Angular Momentum in the Yaw Direction Based on Human Running," in \textit{ Proc. IEEE International Conference on Robotics and Automation (ICRA)}, 2017. 

\bibitem{ITUAtmAte} Radiocommunication Sector of International Telecommunication Union (ITU), ``Recommendation ITU-R P.676-10: Attenuation by atmospheric gases- P Series Radiowave propagation," Tech. Rep. ITU-R P.676-10, 2013. 

\bibitem{Part15++} R. Singh, D. Sicker, ``Part 15++: An Enhanced ID-based Approach for Etiquette and Enforcement Management in Unlicensed Band," \textit{IEEE Transactions on Cognitive Communications and Networking (TCCN)}, 2019. DOI: 10.1109/TCCN.2019.2915662

\bibitem{OnlineML} 	T. M. Mitchell, \textit{Machine Learning}. New York: McGraw-Hill, 1997.



\end{thebibliography}
\end{document}